\documentstyle[12pt]{article}
\begin{document}
\begin{center}
{\large\bf The Exact Solution of the Cauchy Problem for a generalized}

{\large\bf "linear" vectorial Fokker-Planck Equation - Algebraic Approach}


\vspace*{0.2truein}

{A. A. Donkov$^1$, A. D. Donkov$^2$ and E. I. Grancharova$^3$}
\vspace*{0.15truein}

$^1$ {\footnotesize\it Dept. of Physics, University of Wisconsin, 1150 Univ-Ave,
Madison, WI-53706, USA } \\
e-mail: donkov@phys-next1.physics.wisc

$^{2}$ 
{\footnotesize\it Dept. of Physics, University of Sofia,
5 J. Bourchier Blvd.
 Sofia 1164, BULGARIA}     \\
e-mail: donkov@phys.uni-sofia.bg\\
{\footnotesize and \it Bogoliubov Laboratory of Theoretical Physics,
JINR, Dubna, Moscow Region, RUSSIA}\\
e-mail: donkov@thsun1.jinr.ru

$^{3}$
{\footnotesize\it Dept. of Physics, University of Sofia,
5 J. Bourchier Blvd.
 Sofia 1164, BULGARIA}     \\

e-mail: granch@phys.uni-sofia.bg
\vspace*{0.2truein}
\end{center}

\begin{abstract}
The exact solution of the Cauchy problem for a generalized 
"linear" vectorial Fokker-Planck equation
is
found using the disentangling techniques of R. Feynman and algebraic
(operational) methods. This approach may be considered as a generalization of
the Masuo Suzuki's method for solving the 1-dimensional linear Fokker-Planck
equation.
\end{abstract}

\section{Introduction}

The Fokker-Planck equations (FPE), the one-dimensional FPE
\begin{equation}
\frac{\partial W}{\partial t} = - \frac{\partial}{\partial x}
\left[ a(t, x) W \right]
+ \frac{\partial^2}{\partial x^2}
\left[ D(t, x) W(t,x) \right],
\;\;\;\;\;\;\;\;\;\;\;\;\;
t \geq 0,
\;\;\;
x \in {\rm R},
\label{ScFPE}
\end{equation}
\noindent
and the "vectorial" FPE
\begin{equation}
\frac{\partial w}{\partial t} = - \nabla \cdot
                \left[ {\bf a}(t,{ \bf x}) w \right]
+ \nabla \nabla : \left\{ \hat D (t,{ \bf x}) w(t,{\bf x}) \right\},
\;\;\;\;\;\;\;\;\;\;\;\;\;
t \geq 0,
\;\;\;
\bf x \in {\rm R}^n,
\label{VekFPE}
\end{equation}
where $ {\bf a} (t, {\bf x}) = \left( a_1(t,{ \bf x}), a_2(t, {\bf x}), \ldots, a_n(t,
{\bf x})
                                        \right)^{T} $ is the "drift vector",
$\hat D (t, \bf x) $
is a symmetric non-negative definite "diffusion" tensor field of II
rank, and
$ \nabla \nabla : \hat D = \frac{ \partial^2 D_{ij } }
                                { \partial x_{i} \partial x_{j} }
$ (Einstein summation convention accepted),
are widely used~\cite{Fokker}$-\!\!$~\cite{Drozdov} as a tool in
modelling various processes in many areas of the theoretical and
mathematical physics, chemistry and biology as well as in the pure and
applied mathematics and in engineering:
{\it the nonequilibrium statistical mechanics} (in particular in the
theory of Brownian motion and similar phenomena: random walks, the
fluctuations of the liquid surfaces, the local density fluctuations in
fluids and solids, the fluctuations of currents, etc); {\it the metrology}
 ( Josephson voltage standards); {\it the laser physics; the
turbulence theory; the cellular behaviour;
the neurophysiology; the population genetics; the mathematical theory and applications of
 the stochastic
processes},---
to mention only a few of them.

Because of its importance
there have been many attempts to solve FPE exactly or approximately
( for a review see [4, 6 - 11, 14, 19]).
 Among the recent investigations on this problem noteworthy for us  
is the method of M. Suzuki~\cite{Suzuki83}.

In this paper we find the exact solution of following Cauchy
problem:

\begin{equation}
\frac{\partial u}{\partial t} = a_1(t) u(t,{\bf x}) +
{{\bf a}_2}(t) \cdot \nabla u + a_3(t) {\bf x} \cdot \nabla u
+ \hat a_4(t) : \nabla \nabla u,
\;\;\;
u(0,{\bf x})=\phi({\bf x}),
\label{TD}
\end{equation}
where
 $\hat a_4 (t) $ is a symmetric non-negative definite tensor function of
second rank of the scalar parameter $t$.

It is easy to see that the Eq.(\ref{TD}) is
connected with the "linear" vectorial FPE~(\ref{VekFPE}) with a linear
in ${\bf x}$ "drift vector"
$ {\bf a} (t, {\bf x}) = {\bf b}_1 + b_2 {\bf x} $
\quad and an independent of ${\bf x}$ diffusion tensor $\hat D$. ( Here
{\bf b}$_1$, {b}$_2$ and ${\hat D}$
are
functions of $t$.)  
Therefore the Eq.~(\ref{TD}) is a  slight generalization of the
"linear" vectorial FPE (\ref{VekFPE}) with
t-dependent coefficients.

In the paper~\cite{Donkov} the "isotropic" problems

\begin{equation}
\frac{\partial u}{\partial t} = a_1 u(t,{\bf x}) +
{{\bf a}_2} \cdot \nabla u + a_3 {\bf x} \cdot \nabla u + a_4 \Delta u,
\;\;\;\;\;\;\;\;\;\;\;\;
u(0,{\bf x})=\phi({\bf x})
\end{equation}

\noindent
and
\begin{equation}
\frac{\partial u}{\partial t} = a_1(t) u(t,{\bf x}) +
{{\bf a}_2}(t) \cdot \nabla u + a_3(t) {\bf x} \cdot \nabla u
+ a_4(t) \Delta u,
\;\;
u(0,{\bf x})=\phi({\bf x})
\end{equation}
have been exactly solved  ( here $a_4$ and
$a_4(t)$ are arbitrary non-negative constant and function of $t$
respectively).

In the paper \cite {Donkov2} we have found the exact solutions of the
following Cauchy problems:
\begin{equation}
\frac{\partial u}{\partial t} = a_1 u(t,{\bf x}) +
{{\bf a}_2} \cdot \nabla u + a_3 {\bf x} \cdot \nabla u +
\hat a_4 :\nabla \nabla u,
\;\;\;\;
u(0,{\bf x}) = \phi({\bf x})
\label {e1}
\end{equation}
and
\begin{equation}
\frac{\partial u}{\partial t} = a_1(t) u(t,{\bf x}) +
{{\bf a}_2}(t) \cdot \nabla u + a_3(t) {\bf x} \cdot \nabla u +
 a_4(t) \hat a :\nabla \nabla u,
\;\;\;\;
u(0,{\bf x}) = \phi({\bf x})
\label {e2}
\end{equation}
where $\hat a_4$ and $\hat a$ are symmetric non-negative definite tensors of
second rank and $a_4(t)$ is a scalar function; $ a_4(t) > 0 $ . 
( It is obvious
that the
problem~(\ref{TD}) is more general than the problem~(\ref{e2}) :
in~(\ref{TD}),
$\hat a_4(t)$ is arbitrary symmetric non-negative definite tensor function
of second rank, while in~(\ref{e2}) $\hat a_4(t)$ has a special form:
 $\hat a_4(t)= a_4(t) \hat a $. )

Our method may be regarded as a combination of the
disentangling techniques of R. Feynman \cite{Feynman} with the
operational methods 
developed in the functional
analysis and in particular in the theory of pseudodifferential
equations with partial derivatives
~\cite{Maslov}$-\!\!$~\cite{Taylor}.
As we have emphasized in~\cite{Donkov} and~\cite{Donkov2} this approach
is an extension
and generalization of the M. Suzuki's method \cite{Suzuki83} for solving
the one-dimensional linear FPE (\ref{ScFPE}).

\section{Exact Solution of the Cauchy Problem (\ref{TD})}

In view of the t-dependence of the coefficients in the Eq.~(\ref{TD}),
formally we have for the solution of the initial value problem~(\ref{TD})
an ordered exponential
\begin{equation}
u(t,{\bf x}) = \left( \exp_+ \int_{0}^{t}
\left[ a_1(s) +{\bf a}_2(s)\cdot \nabla + a_3(s){\bf x}\cdot \nabla +
\hat a_4(s) : \nabla \nabla \right] {\rm d}s \right) \phi({\bf x}),
\label{form}
\end{equation}
where
$$ 
\exp_+ \int_{0}^{t}\hat C(s){\rm d}s
\equiv T\!\!-\!\exp \int_{0}^{t} \hat C (s) {\rm d}s
$$
\begin{equation}
      = \hat 1 +\lim_{k\to\infty}
\sum_{n=1}^k \int_{0}^{t}{\rm d}t_1 \int_{0}^{t_1}{\rm d}t_2 \dots
\int_{0}^{t_{n-1}} {\rm d} t_{n} \hat C(t_1) \hat C(t_2)\dots \hat C(t_{n}).
\label{texp}
\end{equation}
If we introduce the operators
\begin{equation}
\hat A(t) = {{\bf a}_2}(t) \cdot \nabla + a_3(t) {\bf x} \cdot \nabla
\;\;\;\;\mbox{and}\;\;\;\;
\hat B(t) = \hat a_4(t) : \nabla \nabla,
\label{oper}
\end{equation}
we may write~(\ref{form}) in the form
\begin{equation}
u(t,{\bf x}) ={\rm e}^{\int_{0}^{t} a_1(s){\rm d}s}
\left(\exp_+ \int_{0}^{t}\left[ \hat A(s)+\hat B(s)\right] {\rm d}s \right)
\phi({\bf x}) ,
\label{Form}
\end{equation}
as the first term in the exponent commutes with all others.

To proceed with the pseudodifferential operator in Eq.~(\ref{Form})
we shall use the theorem of M.Suzuki~\cite{Suzuki83} :

If
$$
\left[ \hat A(t) ,\hat B(t) \right] = \alpha(t,s) \hat B(s) ,
$$
then
\begin{equation}
\exp_+ \int_{0}^{t} \left[ \hat A(s) + \hat B(s) \right] {\rm d}s =
\left( \exp_+ \int_{0}^{t} \hat A(s) {\rm d}s \right)
\left( \exp_+ \int_{0}^{t} \hat B(s) {\rm e}^{ -\int_{0}^{s}
\alpha(u,s) {\rm d}u }{\rm d} s \right).
\label{Th}
\end{equation}

In our case we have
$$
\left[ \hat A(s), \hat B(s') \right]
\equiv
\left[\; {\bf a}_2(s) \cdot \nabla + a_3(s) {\bf x} \cdot \nabla \;\;,\;\;
\hat a_4(s') : \nabla \nabla \;\right]
$$
\begin{equation}
=
- 2 a_3(s) \hat a_4(s') :\nabla \nabla \equiv  - 2 a_3(s) \hat B(s') .
\label{Com}
\end{equation}
Therefore from~(\ref{Th}) we obtain
\begin{equation}
\exp_+ \int_{0}^{t} \left[ \hat A(s) + \hat B(s) \right] {\rm d}s =
\left( \exp_+ \int_{0}^{t} \hat A(s) {\rm d}s \right)
\left( \exp_+ \int_{0}^{t} \hat B(s) {\rm e}^{ 2\int_{0}^{s}
 a_3(u) {\rm d}u }{\rm d} s \right)    .
\label{ThN}
\end{equation}
The linearity  of the integral and the explicit form of $\hat A $
(see Eq.~(\ref{oper})) permit to write the first factor in~(\ref{ThN})
in terms of usual, not ordered, operator valued exponent
\begin{equation}
\exp_+ \int_{0}^{t} \hat A(s) {\rm d}s \equiv
\exp_+ \int_{0}^{t} \left[ {\bf a}_2(s) \cdot \nabla + a_3(s) {\bf x}
\cdot \nabla \right] {\rm d}s =
{\rm e}^{\vec \alpha_2(t) \cdot \nabla + \alpha_3(t){\bf x} \cdot \nabla}.
\label{hatA}
\end{equation}

\noindent 
For convenience we introduce the following notations:
\begin{equation}
\alpha_1(t)=\int_{0}^{t} a_1(s) {\rm d}s, \;\;
{\vec \alpha}_2(t)=\int_{0}^{t} {\bf a}_2(s) {\rm d}s, \;\;
\alpha_3(t)=\int_{0}^{t} a_3(s) {\rm d}s. \;\;
\label{alpha}
\end{equation}
Consequently (from now on "$'$" means $\frac{d}{dt}$)
$$
\alpha'_1(t)=a_1(t), \;\;
{\vec \alpha}'_2(t)={\bf a}_2(t), \;\;
\alpha'_3(t)=a_3(t), \;\;
$$
\begin{equation}
\alpha_1(0)=0, \;\;
{\vec \alpha}_2(0)={\bf 0}, \;\;
\alpha_3(0)=0. \;\;
\end{equation}

 Thus we obtain from the Eq.~(\ref{Form})
\begin{equation}
u(t, {\bf x})=
{\rm e}^{\alpha_1(t)}
{\rm e}^{[{\vec \alpha}_2(t) + \alpha_3(t){\bf x}] \cdot
\nabla}
\left( \exp_+ \int_{0}^{t} \hat a_4(s) {\rm e}^{2 \alpha_3(s)} : \nabla \nabla
{\rm d}s \right) \phi ({\bf x}).
\label{u}
\end{equation}
Finally using the formulae (see~\cite{Donkov} and~\cite{Donkov2}) \\
\vspace{0.1truein}

\noindent
$\left[ exp_+ \left( \int_{0}^{t} \hat \Psi(s):\nabla \nabla{\rm d}s \right)\right]$
$ \phi ({\bf x})$ 
\begin{equation}
=\frac{ 1 }
     { \sqrt{\det (4 \pi \hat \tau(t))} }
\int\limits_{\rm R^n}
\left\{       \exp
       \left[
               - ({\bf x-y} )
\cdot
\frac{\hat \tau^{-1}(t)}{4}
\cdot
                ( {\bf x-y} )
        \right]
          \right\}
\phi({\bf y})
{\rm d}y,
\label{Phi}
\end{equation}
where
$$
{\rm d}y = {\rm d}y_1 {\rm d}y_2 \dots {\rm d}y_{n} ,
\;\;\;\;\;\;\;\;
\hat \tau(t) = \int_{0}^{t} \hat \Psi(s) {\rm d}s
$$
and

\begin{equation}
{\rm e}^{ {\vec \alpha}_2(t) \cdot \nabla
+ \alpha_3(t) {\bf x} \cdot \nabla }
g({\bf x}) =
g\left(
      {\bf x} {\rm e}^{\alpha_3(t)}
      + \int_{0}^{t}
            {\bf a}_2(s)
            {\rm e}^{\alpha_3(s)}
         {\rm d}s
\right)
\equiv
g( {\bf z} ),
\label{ge}
\end{equation}
we find from the Eq.~(\ref{u})
the following expression for the exact solution of the Cauchy
problem~(\ref{TD})
$\left( \hat \Psi(s) = \hat a_4(s) \exp[2a_3(s)] \right)$ :
\begin{equation}
u(t, {\bf x}) =
\frac{ {\rm e}^{\alpha_1(t)}}
     { \sqrt{\det (4 \pi \hat \tau(t))} }
\int\limits_{\rm R^n}
\left\{       \exp
       \left[
               - ({\bf z-y} )
\cdot
\frac{\hat \tau^{-1}(t)}{4}
\cdot
                ( {\bf z-y} )
        \right]
          \right\}
\phi({\bf y})
{\rm d}y,
\label {end}
\end{equation}
where
$$
\hat \tau(t) = \int_{0}^{t} \hat a_4(s)
{\rm e}^{ 2 \alpha_3(s)}{\rm d}s
$$
is a symmetric non-negative definite second rank tensor function of $t$
, ${\rm d}y = {\rm d}y_1  \dots {\rm d}y_{n} $
and ${\bf z}$ is defined in~(\ref{ge}).

Substituting the expression~(\ref{end}) in the Eq.~(\ref{TD}) we see
immediately that the function
$u(t, {\bf x})$ is a solution of the problem~(\ref{TD}), and, according to the
Cauchy theorem, it is the only classical solution of this problem.

\section{Concluding remarks}

\begin{itemize}
\item The exact solutions of the Cauchy problem (\ref{TD})
is obtained using the algebraic method we have described.
\item When
$\hat a_4 (t)$  
is scalar: 
$\hat a_4(t)= a_4(t) \hat 1$  
( in this case
$\hat a_4 : \nabla \nabla = a_4 \Delta$ )
the "anisotropic" problem~
(\ref{TD}) turns to the
"isotropic" one, with the exact solution found in~\cite{Donkov}. It is
easy to check that the solution~(\ref{end}) 
turns to
the solution obtained in~\cite{Donkov}
(there is an error in~\cite{Donkov}: the sign before {\bf a}$_2$ in the
Eqs. (17) and (34) there, should be (+)).
\item In the case $ \hat a_4(t) = a_4(t) \hat a $ the Cauchy problem~(\ref{TD})
reduces to the problem~(\ref{e2}) treated in~\cite{Donkov2}.
In this case the solution~(\ref{end}) turns to the solution obtained
in~\cite{Donkov2}.
\item For different choices of the coeficients $a_j$
and ${\bf a}_2$ the Eq.~
(\ref{TD}) may be regarded also as a set of different diffusion equations.
Therefore from the formula~(\ref{end}) 
we obtain the exact
solutions of the Cauchy problems for this set of diffusion equations.
\end{itemize}


\end{document}